\documentclass[%
 reprint,twocolumn,floatfix,
 amsmath,amssymb,
 aps,
 prx,
 showpacs,
longbibliography,
superscriptaddress
]{revtex4-2}

\usepackage{graphicx}
\usepackage{color}
\usepackage{comment}
\usepackage{url}
\usepackage[normalem]{ulem}
\usepackage[version=3]{mhchem} 

\newcommand*{\rr}{{\mathbf r}} 
\newcommand*{\Ey}{E_{\tn Y}} 
\newcommand*{\tey}{\tilde{E}_{\tn Y}} 
\newcommand*{\bR}{\bar{R}} 
\newcommand*{\br}{\bar{r}} 
\newcommand*{\bRRR}{\overline{R^3}}
\newcommand{\tn}[1]{\textnormal{#1}}  

\graphicspath{{figures/}}

\begin{document}

\title{Phase Separation and Ripening in a Viscoelastic Gel}

\author{Tine Curk}
\email{curk@northwestern.edu}
\affiliation{Department of Materials Science \& Engineering,
  Northwestern University, Evanston, Illinois 60208, USA}

\author{Erik Luijten}
\email{luijten@northwestern.edu}
\affiliation{Department of Materials Science \& Engineering,
  Northwestern University, Evanston, Illinois 60208, USA}
\affiliation{Departments of Engineering
  Sciences \& Applied Mathematics, Chemistry and Physics \& Astronomy,
  Northwestern University, Evanston, Illinois 60208, USA}

\date{\today}

\begin{abstract}
  The process of phase separation in elastic solids and viscous fluids is of
  fundamental importance to the stability and function of soft materials.  We
  explore the dynamics of phase separation and domain growth in a viscoelastic
  material such as a polymer gel. Using analytical theory and Monte Carlo
  simulations we report a new domain growth regime, in which the domain size
  increases algebraically
  with a ripening exponent~$\alpha$
  that depends on the viscoelastic properties of the material. For a
  prototypical Maxwell material, we obtain $\alpha=1$, which is markedly
  different from the well-known Ostwald ripening process with $\alpha=1/3$.
  We generalize our theory to systems with arbitrary power-law relaxation
  behavior and discuss our findings in the context of the long-term stability
  of materials as well as recent experimental results on phase separation in
  cross-linked networks and cytoskeleton.
\end{abstract}

\maketitle

\section{Introduction}

Phase separation is a universal phenomenon and a fundamental concept in
physics, chemistry and materials science. Whereas its outcome is described by
equilibrium thermodynamics, it is equally important to understand the dynamics
of the separation process. This is particularly relevant in solid, soft and
biological systems where large kinetic barriers or active processes often
prevent the system from reaching its equilibrium state.  Upon formation of a
new phase, domains typically grow via coalescence or via
surface-tension-driven dissolution and redeposition known as Ostwald
ripening~\cite{RatkeVoorhees2002}.  However, elastic effects can markedly
alter the dynamics of these processes.  As shown in the early days of
solid-state physics, nucleation and precipitation in crystalline solids cause
substantial elastic stresses, which, in turn, control the shape and growth of
precipitate particles and the macroscopic mechanical properties of the
material~\cite{Nabarro1940}.  Recent experiments on droplet formation within
disordered polymer networks have shown that elastic stress can either fully
inhibit liquid--liquid phase separation or arrest the coarsening
process~\cite{Style2018,Kim2020}. The latter can lead to migration of droplets
in stiffness gradients, ``elastic ripening''~\cite{Dufresne2020}, which is
supported by theoretical
studies~\cite{Kothari2020,Vidal-Henriquez2020,Wei2020,Vidal-Henriquez2021,Ronceray2022}.

However, the situation is completely different in materials that exhibit
viscoelastic relaxation or creep, which can irreversibly reduce elastic stress
and thus enable further domain coarsening.  Viscoelastic effects in phase
separation have been extensively studied in spinodal
decomposition~\cite{Siggia1979}, polymer solutions~\cite{Tanaka2000},
viscoelastic fluids~\cite{Tabuteau2009} and glass-forming
melts~\cite{Pascova1990,Schmelzer1990,Schmelzer1990c,Schmelzer2003}, as well
as in protein and colloidal suspensions~\cite{Tanaka2005,Tateno2021}.
However, to the best of our knowledge phase separation of a minority phase of
small molecules inside a polymer matrix has only been studied on short time
scales~\cite{Pascova1990,Style2018}, where coarsening is arrested by the
elastic stress. In contrast, the long-time domain growth of nuclei or droplets
is governed by the interplay between elastic forces and viscous flow. The
associated coarsening laws are currently unknown. Consequently, a variety of
material aging processes currently cannot be predicted.

This work was originally motivated by metal soap formation in oil paintings,
an aging process believed to be driven by phase
separation~\cite{Casadio2019book}.  However, the relevance of our findings
goes well beyond this topic, as it applies to various branches of materials
science and is believed to play a crucial role in biological systems.  Phase
transitions involving biomolecular liquids have been established as
fundamental drivers of intracellular
organization~\cite{Feric2016,Alberti2020}, and viscoelastic relaxation can
enable cells to flexibly modulate mechanical properties of membrane-less
organelles~\cite{Weber2017,Jawerth2020,Michieletto2022}.  The realization that coupling
between condensation and network elasticity may play a role in cellular
physiology has led to renewed interest in phase separation in
cells~\cite{Shin2018,Lee2021} and in synthetic
materials~\cite{Style2018,Dufresne2020}. 

Here, we employ analytical theory and particle-based Monte Carlo simulations
to investigate the process of phase separation and surface-tension-driven
coarsening within a viscoelastic medium. Focusing on a prototypical system
where a minority phase forms spherical nuclei or droplets within a
viscoelastic majority phase, we attain a fundamental and quantitative
understanding of the coarsening process on both short and long time scales.

\section{Phase separation in elastic media}

\begin{figure}
\centering
\includegraphics[width=0.48\textwidth]{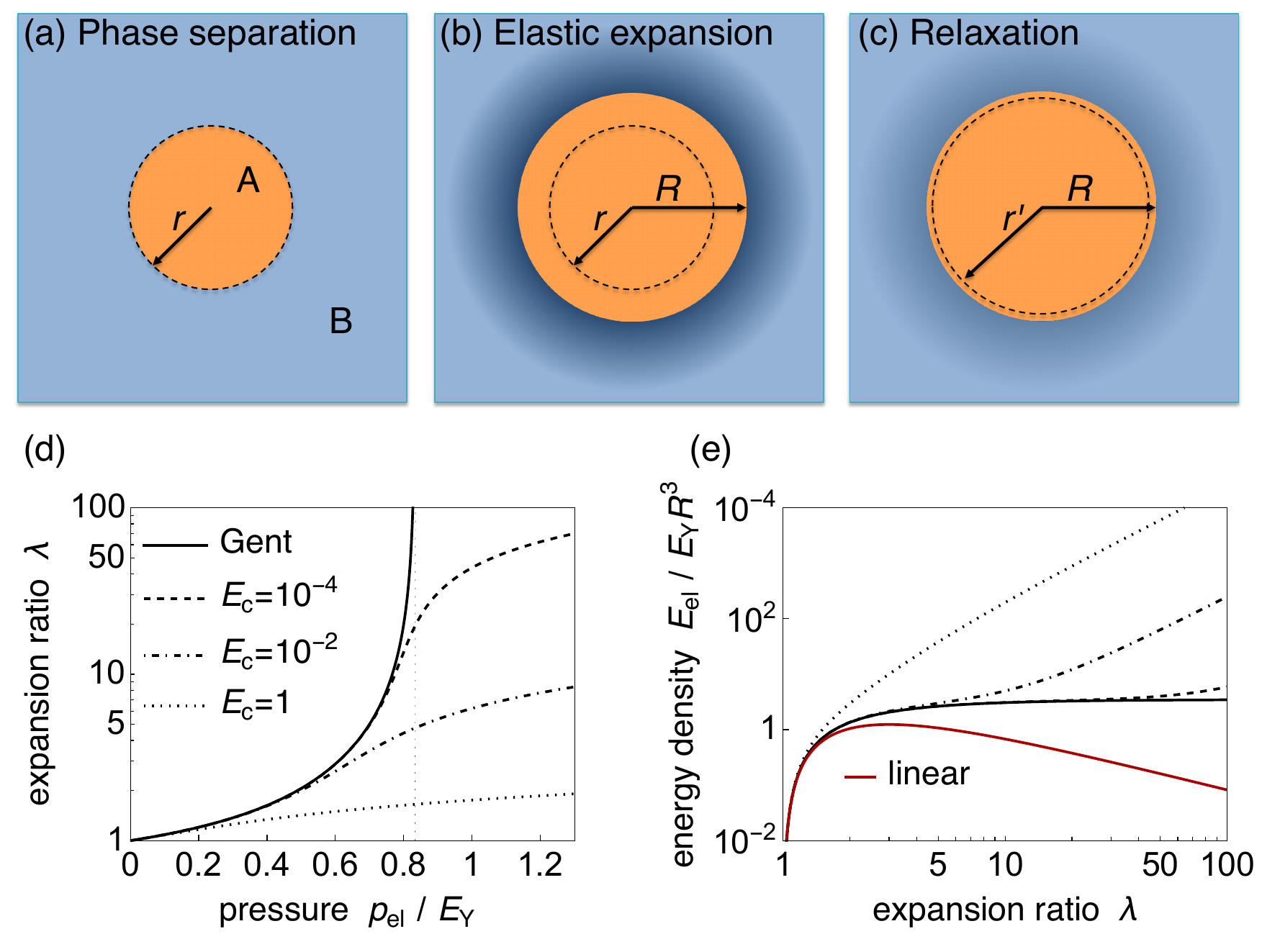}
\caption{Growth of a droplet in a viscoelastic medium. (a)~Schematic of a
  minority phase~A (orange) in an elastic majority phase~B (blue), with $r$
  the unstressed cavity size. (b)~Growth of the droplet to size $R=\lambda r$
  causes elastic stresses in phase~B (indicated by a darker shade of
  blue). (c)~Viscoelastic relaxation expands the cavity to size~$r'$, thus
  reducing the stress and enabling further growth of phase~A\@. (d)~Droplet size
  dependence on pressure for a Gent material [Eq.~\eqref{eq:pel}] (solid line)
  and for a material with additional nonlinear elasticity proportional
  to~$E_{\tn c}$ (dashed, dot-dashed and dotted lines). The thin dotted
  vertical line indicates the elastic cavitation pressure $p_{\tn c}=5\Ey/6$.
  (e)~Elastic energy density corresponding to examples in~(d) and comparison
  with the linear elastic case (solid red line)~\cite{LLelasticity1959book}.}
\label{fig:scheme}
\end{figure}

To set the stage for our investigation of viscoelasticity, we first briefly
introduce elastic effects in phase separation and coarsening.  We limit our
analysis to the dilute regime, so that many-body elastic effects can be
neglected and each droplet or precipitate can be considered an isolated
spherical inclusion in an isotropic elastic medium
(Fig.~\ref{fig:scheme}a). The assumption of spherical symmetry makes this
problem analytically tractable.  The initial cavity size~$r_0$ is determined
by the pore size in the material or the crosslink distance. For example,
$r_0^3 \sim k_{\tn B}T/\Ey$ for a flexible gel with Young's modulus
$\Ey$~\cite{rubinstein03}, with $T$ the absolute temperature and $k_{\tn B}$
the Boltzmann constant. An inclusion of size~$R$ growing beyond the cavity
size will produce an elastic stress whose radial component at the cavity
interface~$\sigma$ is balanced by the excess pressure within the inclusion,
$p_{\tn{el}} = \sigma$.  (Fig.~\ref{fig:scheme}b). Viscoelastic relaxation can
increase the cavity size, enabling further growth of the inclusion
(Fig.~\ref{fig:scheme}c).

To describe the elastic expansion of the cavity we use the Gent model for the
elastic pressure of a spherical inclusion in a rubber-like nonlinear
solid~\cite{Gent1991,Gent1996}, a good approximation for cross-linked
networks~\cite{Crosby2007,Crosby2016,Style2018,Dufresne2020},
\begin{equation}
  p_{\tn{el}}^{\tn{Gent}}(\lambda) = \frac{\Ey}{6} \left( 5- \frac{4}{\lambda}
    - \frac{1}{\lambda^{4}} \right)\;,
\label{eq:pel}
\end{equation}
with $\lambda \equiv R/r$ the expansion ratio and $r$ the radius of the
unstressed cavity (Fig.~\ref{fig:scheme}).  At pressures exceeding the
cavitation pressure $p_{\tn c}=5\Ey/6$, elastic forces are unable to contain
the inclusion and its size~$R$ increases without bounds
(Fig.~\ref{fig:scheme}d). However, it is known that additional
strain-stiffening effects can limit the size at very
high~$\lambda$~\cite{Crosby2007,Crosby2016,Style2018,Dufresne2020}. We account
for this through an additional term
$p_{\tn{el}}^{\tn{lim}} = E_{\tn c} \Ey (\lambda-1)^2$, with dimensionless
prefactor $E_{\tn c} > 0$, that limits the expansion ratio
to~$\lambda_{\tn c}\sim E_{\tn c}^{-0.5}$~\footnote{The exact expression for
  $p_{\tn{el}}^{\tn{lim}}$ is not of crucial importance. Any functional form
  that limits the expansion ratio,
  $p_{\tn{el}}^{\tn{lim}} \sim \lambda^{\kappa}$ as $\lambda \to \infty$, with
  $\kappa > 0$, could be used.}. We refer to this as the extended Gent model
and note that a different strain-stiffening description such as the
Mooney--Rivlin model~\cite{Kothari2020} or the Gent model with a set maximum
strain~\cite{Wei2020}, would not affect our observations.  On the other hand,
at small deformations ($\lambda\sim1$) Eq.~\eqref{eq:pel} reduces to
$p_{\tn{e}} = \frac{4}{3}\Ey (\lambda-1)$, which corresponds to a standard
result of linear elasticity,
$p_{\tn{el}} = 2\Ey (\lambda-1)/(1+\nu)$~\cite{LLelasticity1959book} with
Poisson ratio $\nu=1/2$. Thus, at small deformations our findings apply to any
isotropic, linear elastic material.

We note that when the size of inclusions is allowed to equilibrate through the
exchange of mass via an evaporation--condensation process, the equilibrium
size~$R$ can be calculated using classical nucleation theory augmented with an
elastic energy term $E_{\tn{el}}(\lambda)$. For a spherically symmetric
system, this elastic energy equals the reversible pressure work performed by
expanding the cavity from its unstressed radius~$r$ to $R=\lambda r$,
$E_{\tn{el}}(\lambda) = 4 \pi \int_{r}^{R} r'^2 p_{\tn{el}}(r'/r) dr' \;$.
For the Gent model, the full analytical expression is provided in the Methods
section [Eq.~\eqref{eq:Eel}].  Surface tension favors large~$R$, while elastic
pressure favors small~$R$.  Competition between surface tension and elastic
stress results in a well-defined equilibrium
size~$R$~\cite{Style2018,Dufresne2020}.

\section{Simulations of viscoelastic ripening}
\label{sec:simu}

Having described the (reversible) elastic response we emphasize that
understanding the phase separation process requires a model of irreversible
deformations.  The simplest description of a material that is elastic on short
time scales but can flow on long time scales is a Maxwell material, which is
characterized by a single relaxation time~$\tau_{\tn r}$. For example,
supramolecular networks~\cite{Tabuteau2009} and
vitrimers~\cite{Denissen2017,Kalow2020}, but also covalently cross-linked
polymers~\cite{Capiel2020}, exhibit Maxwell-type viscoelastic relaxation on
sufficiently long time scales. Moreover, both crystalline and amorphous
materials are generally subject to diffusion creep that follows a single
time scale relaxation process. In contrast, biopolymer
networks~\cite{Mulla2019}, protein condensates~\cite{Jawerth2020} and glassy
materials~\cite{Sollich1997} exhibit aging and relaxation over multiple
time scales and are instead described by a Maxwell glass model.  For example,
the strain~$\varepsilon$ of cytoskeleton under stress~$\sigma$
increases as a power law, $\varepsilon \sim \sigma t^{a}$, with
exponent~$a \approx 0.5$ found both theoretically~\cite{Broedersz2010} and
experimentally~\cite{Desprat2005}.

\begin{figure}
\centering
\includegraphics[width=0.45\textwidth]{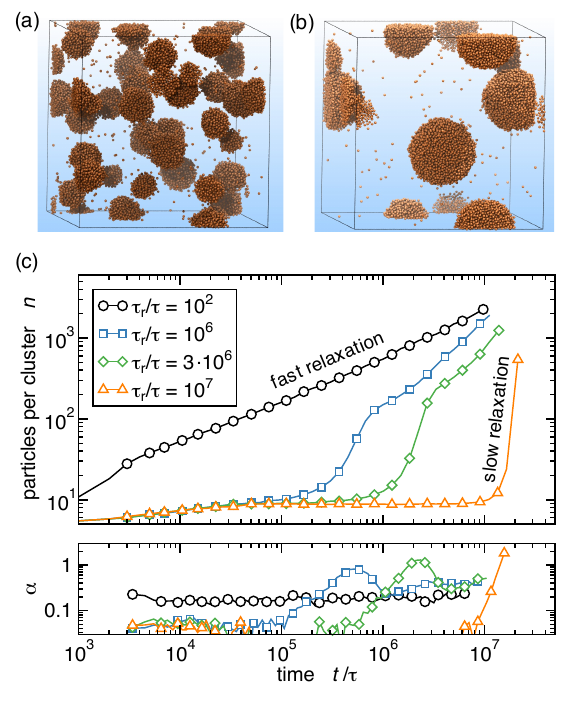}
\caption{Growth of condensates in a viscoelastic medium with relaxation
  time~$\tau_{\tn r}$ as determined by Monte Carlo simulations.
  (a,b)~Snapshots at times $t=8\cdot 10^5\tau$~(a) and $t= 10^7\tau$~(b) for
  $\tau_{\tn r}= 10^6 \tau$. Individual spherical particles with
  diameter~$d_{\tn {LJ}}$ are shown as orange spheres and the box outline
  indicates periodic boundaries. (c)~Top panel: Growth in the average number
  of particles per cluster~$n$ for various $\tau_{\tn r}$. Bottom panel:
  Ripening exponent~$\alpha$ [(Eq.~\eqref{eq:alpha})], which equals $1/3$ of
  the slope in the top panel, since $n \sim \bR^{3}$. Elastic parameters:
  $\Ey=k_{\tn B}T/d_{\tn{LJ}}^3$ and $E_{\tn c}=1$.}
\label{fig:MC}
\end{figure}

Coarsening processes are usually modeled using continuum mean-field
theories~\cite{RatkeVoorhees2002}, which very accurately predict growth laws
in the large-domain and the long-time limits.  We will develop such a theory
in the next section. Modeling of small clusters on short time scales, however,
requires a more detailed model that is able to accurately capture the
interplay between nucleation, phase separation and viscoelastic relaxation. To
this end, we first employ a Lennard-Jones (LJ) fluid as a prototypical
particle-based model and couple it to a viscoelastic background medium with
embedded spherical cavities. The configuration of this system is evolved using
single-particle Monte Carlo displacement moves, modeling diffusive dynamics of
individual particles.  Nucleation and growth of liquid droplets within
cavities causes stress and increases elastic energy.  This modifies the rate
at which individual particles are added to or removed from a droplet, which is
accounted for through the Monte Carlo acceptance probabilities.  In addition,
the elastic stress leads to irreversible expansion of the cavities depending
on the viscoelastic properties of the medium (see \emph{Methods}).

The viscoelastic medium is described by the upper-convected Maxwell model~\cite{Bird1987}. 
In the small stress limit, this model reduces to the Maxwell model (see Supporting Information) and we use it to enable analytical predictions.
For the ripening process
considered, large stress can only occur when the clusters are small
(stress scales as $\sigma \sim 1/R$, see Eq.~\eqref{eq:sigmag} below),
in which case the Maxwell model yields an upper bound for the coarsening
rate. Thus, we employ the Maxwell model and note that this approximation does
not affect the prediction of the coarsening exponents, which are defined in
the large-$R$ limit.

For a Maxwell material at constant strain, the elastic stress~$\sigma$ decays
exponentially, $\partial \sigma / \partial t = -\sigma / \tau_{\tn r}$, on a
time scale $\tau_{\tn r}= \eta / \Ey$, with $\eta$ the material
viscosity. To avoid artifacts due to non-linear elasticity, we use a
generalized form based on the reduction of the elastic energy~$E_{\tn{el}}$,
\begin{equation}
  \left(\frac{\partial E_{\tn{el}}}{\partial t}\right)_R = -2\frac{E_{\tn{el}}}{\tau_{\tn r}}
  \;,
\label{eq:Erelax}
\end{equation}
which in the linear elasticity regime corresponds exactly to the Maxwell
model. Note that $E_{\tn{el}}=E_{\tn{el}}(R,r)$ is a state function and its
time dependence arises solely through variation of the dynamical variables $R$
and~$r$. Factorization $E_{\tn{el}}$ [Eqs.\ \eqref{eq:Eels}
and~\eqref{eq:Eelchain} in \emph{Methods}] and insertion into
Eq.~\eqref{eq:Erelax} leads to the expansion rate
\begin{equation}
\frac{\tn{d} r}{\tn{d} t} = \frac{2r}{\tau_{\tn r}} \left[\left(\frac{\partial \ln [E_{\tn{el}}/(E_{\tn Y}r^3)]}{\partial \ln \lambda}\right)_R-3 \right]^{-1}\;.
\label{eq:dR0dt}
\end{equation}

The Maxwell model is valid for small strain rates,
$\tn{d} r / \tn{d} t \ll r / \tau_{\tn r}$. Large strain rates require the
upper-convected Maxwell model, which leads to additional shear
thickening~\cite{Bird1987}, but becomes analytically intractable and has not
been generalized to nonlinear elasticity. For the ripening process
considered, large strain rates can only occur when the clusters are small
(since stress scales as $\sigma \sim 1/R$, cf.\ Eq.\eqref{eq:sigmag} below),
in which case the Maxwell model yields an upper bound for the coarsening
rate. Thus, we employ the Maxwell model and note that this approximation does
not affect the prediction of the coarsening exponents, which are defined in
the large-$R$ limit.

The interplay of condensation and viscoelastic relaxation leads to specific
dynamics of domain coarsening and cluster growth (Fig.~\ref{fig:MC}), which we
characterize via the growth exponent,
\begin{equation}
  \alpha = \frac{\tn{d} \log (\bR)}{\tn{d} \log (t)}\;,
 \label{eq:alpha}
\end{equation}
with $\bR$ the mean cluster radius. Under sufficiently fast relaxation
($\tau_{\tn r} \to 0$), elasticity is irrelevant and the domain growth is
expected to be determined by Ostwald ripening with a power-law exponent
$\alpha=1/3$ in the large-domain limit~\cite{Baldan2002,RatkeVoorhees2002}.
In our particle-based simulations we find $\alpha \approx 0.23$
(Fig.~\ref{fig:MC}c, bottom panel). This underestimation is a finite-size
effect originating from the surface diffusion of individual
particles~\cite{Jeppesen1993}.  However, if the viscoelastic medium relaxes
more slowly, markedly different behavior results. On short time scales,
$t \ll \tau_{\tn r}$, the growth is limited and the evolution of cluster size
is determined by elasticity~\cite{Style2018,Dufresne2020}, whereas on intermediate
time scales, $t \sim \tau_{\tn r}$, viscoelastic relaxation enables expansion
of cavities and continued domain growth. Surprisingly, we discover that in
this \emph{viscoelastic ripening} regime the exponent exceeds the Ostwald
value, $\alpha > 1/3$. Since the simulations are restricted to relatively
small systems and limited time scales, we further explore the nature of this
new regime, which is controlled by viscoelastic relaxation, by means of
perturbation theory.

\section{Viscoelastic ripening theory}

\subsection{Extension of  Landau--Slyozov--Wagner theory}
  
The Landau--Slyozov--Wagner (LSW) theory of Ostwald
ripening~\cite{Lifshitz1961,RatkeVoorhees2002} predicts that the average
radius~$\bR$ of a spherical domains grows as
\begin{equation}
  \frac{\tn{d} \bR}{\tn{d} t} = C c^{\rm sat} \frac{\gamma}{\bR^2}  \;,
\label{eq:ROst}
\end{equation}
where $c^{\rm sat}$ is the monomer concentration in phase~B, $\gamma$ the
surface tension and the prefactor $C=8D/(27 c_{\tn A}^2 k_{\tn B}T)$,
with $D$ the diffusion coefficient and $c_{\tn A}$ the monomer number density
in the condensed phase~A\@. The resulting power law-growth~$\bR\sim t^{1/3}$,
is accompanied by a scale-free size
distribution~$h(R/ \bR)$~\cite{Lifshitz1961,Baldan2002,RatkeVoorhees2002}.

Elastic effects alter this coarsening process in two ways. Firstly, they
change the phase equilibrium between the condensed phase~(A) and the
viscoelastic phase~(B), which is captured by an elasticity-dependent
$c^{\rm sat}$~\cite{Dufresne2020}. Assuming that the monomers in phase B are
dilute and can be described by the ideal chemical potential, we find
\begin{equation}
 c^{\tn{sat}}(p_{\tn{el}}) = c^{\tn{sat}}_{0} e^{p_{\tn{el}}/(c_{\tn A }k_{\tn B}T)} \;,
 \label{eq:csat}
\end{equation}
with the elasticity-free reference value
$c^{\tn{sat}}_{0} = c^{\tn{sat}}(p_{\tn{el}}\to0)$,  Secondly, the elastic energy
alters the coarsening kinetics. We account for this effect using a first-order
perturbation expansion of the LSW theory (see Methods section), leading to the
coarsening rate
\begin{equation}
  \frac{\tn{d} \bR}{\tn{d} t}  =C c^{\tn{sat}}(p_{\tn{el}}) \left[ \frac{\gamma}{\bR^2} -  \frac{1}{4\pi} \frac{\partial [E_{\tn{el}}(\bR,\br)/\bR^3]}{\partial \bR} \right] \;.
\label{eq:prpt}
\end{equation}
The elastic pressure~$p_{\tn{el}}$, the average domain radius~$\bR$ and the
mean cavity radius~$\br$ are all time dependent, while the prefactor~$C$ is
determined by the diffusion constant~$D$ and the density of the condensed
phase~$c_{\tn A}$~\cite{RatkeVoorhees2002}.  We assume the limit in which the
condensed phase~A is incompressible relative to the surrounding viscoelastic
phase~B, and thus both $c_{\tn A}$ and $C$ are considered constant.  The
surface tension~$\gamma$ is positive, while the elastic term in
Eq.~\eqref{eq:prpt} can be either positive or negative depending on the
elastic energy density, and thus can either accelerate or inhibit the
coarsening process (cf.\ Fig.~\ref{fig:scheme}e).  For example, the extended
Gent model leads to growth in $\bR$ until the elastic term balances the
surface tension term, at which point the growth is arrested. Prior to this
arrest, $\bR$ exponentially approaches its equilibrium value, as was predicted
and experimentally verified in glass-forming
melts~\cite{Schmelzer1990}. Moreover, Eq.~\eqref{eq:prpt} predicts that the
equilibrium droplet size decreases when the elastic modulus is increased, as
was observed in crosslinked gels~\cite{Style2018,Dufresne2020}. In the absence
of elasticity, $E_{\tn{el}} = 0$, Eq.~\eqref{eq:prpt} reduces to the LSW
theory, Eq.~\eqref{eq:ROst}.

High droplet concentrations would induce additional interactions between
droplets due to elastic distortion of the material, affecting the above
predictions.  However, departure from the nondilute conditions does not affect
the ripening exponent in the LSW theory but only requires a rescaling of the
prefactor~\cite{RatkeVoorhees2002}. Since elastic deformations are governed by
the same algebraic profiles as the density profiles in the LSW theory, we
argue that elastic interactions between droplets also will not affect the
ripening exponent.

\begin{figure}
\centering
\includegraphics[width=0.45\textwidth]{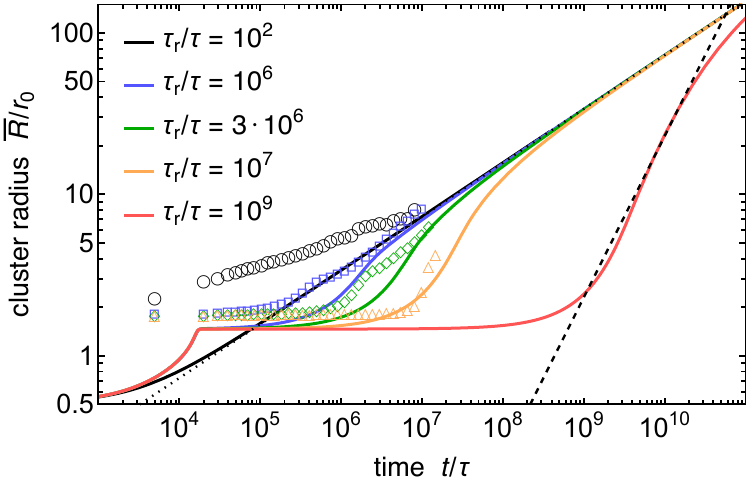}
\caption{Theoretical prediction for cluster growth obtained by numerical
  integration of Eqs.\ \eqref{eq:Erelax} and~\eqref{eq:prpt} (solid lines),
  with corresponding MC simulation results for different relaxation times,
  $\tau_{\tn{r}}/\tau = 10^{2}$ (circles), $\tau_{\tn{r}}/\tau = 10^{6}$
  (squares), $\tau_{\tn{r}}/\tau = 3\cdot10^{6}$ (diamonds) and
  $\tau_{\tn{r}}/\tau = 10^{7}$ (triangles). The dotted black line shows the
  LSW theory prediction, Eq.~\eqref{eq:ROst}, and the dashed line is the small
  deformation approximation to viscoelastic ripening,
  Eq.~\eqref{eq:Rve}. Initial cavity diameter $r_0 = d_{\tn{LJ}}$ and
  $\Ey=k_{\tn B}T/r_0^3$.}
\label{fig:theory}
\end{figure}

To model viscoelastic relaxation, we couple the growth of the cavity [Eq.~\eqref{eq:prpt}] with that of the inclusion [Eq.~\eqref{eq:dR0dt}] and 
integrate the two equations numerically, obtaining a
prediction for domain growth (Fig.~\ref{fig:theory}). Strikingly, our
extension of LSW theory qualitatively captures the observations of
Sec.~\ref{sec:simu}. In particular, we note that the perturbative solution
confirms the existence of the intermediate viscoelastic ripening regime where
the growth exponent, $\alpha \sim 1$, exceeds that of Ostwald ripening
($\alpha \sim 1/3$).

To directly compare the theory to the results of Fig.~\ref{fig:MC}, we
determine the prefactor~$C$ [Eq.~\eqref{eq:prpt}] from the MC simulation
conditions (see Methods section).  Given the first-order nature of the theory,
the semi-quantitative agreement with the MC data is quite remarkable
(Fig.~\ref{fig:theory}). Moreover, deviations at small $\bR$ are to be
expected, since LSW theory is only valid in the large-domain
limit. Conversely, we note that the simulation results for the longest times
exhibit quantitative agreement with the theoretical predictions.
Lastly, we note that the theory illustrates how viscoelastic effects can
change the predicted domain size by orders of magnitude compared to the
standard Ostwald ripening prediction.

\subsection{Analytical theory}

Whereas the perturbative solution resolves the cluster growth process across
all time scales, an analytical theory would describe the viscoelastic ripening
regime without the need for a numerical solution.  Here we derive such a
theory by assuming a small expansion ratio, $\lambda \approx 1$.  In this
limit, the elastic term in Eq.~\eqref{eq:prpt} simplifies to
$\frac{1}{4\pi \bR^2} \frac{\partial E_{\tn{el}}(\bR,\br)}{\partial R }$,
which we recognize as the radial stress at the cavity wall~$\sigma$. Under
elastic arrest, $\tn{d}\bR/\tn{d}t \approx 0$, the stress is thus determined
solely by the surface tension,
\begin{equation}
 \sigma \approx \gamma/\bR \;.
\label{eq:sigmag}
\end{equation} 
We note that this approximation did not require linear elasticity assumptions.
The same relation can be obtained by employing classical nucleation theory and
assuming quasi-equilibrium conditions for a system of spherical clusters at
constant total volume (see Sec.~\ref{sec:ptheory}).  Combining
Eq.~\eqref{eq:Erelax}, which reduces to
$\tn{d}\br/\tn{d}t=(\lambda-1)\br/\tau_{\tn{r}}$, with Eq.~\eqref{eq:sigmag}
and the linear elastic stress,
$\sigma=2 \tey (\lambda-1)$~\cite{LLelasticity1959book}, where
$\tey=\Ey/(1+\nu)$, results in an analytical prediction for the growth rate,
\begin{equation}
  \frac{\tn{d} \bR}{\tn{d} t} = \frac{\gamma} {2\tau_{\tn r} \tey} \;.
\label{eq:Rve}
\end{equation}
This expression is valid for sufficiently large clusters, $R > \gamma/\tey$,
such that strain rates are low and Maxwell model is appropriate. For small
clusters $R \le \gamma/\tey$ Eq.~\eqref{eq:Rve} represents an upper bound of
the growth rate.  Equation~\eqref{eq:Rve} is in excellent agreement with the
numerical integration result in the viscoelastic ripening regime with exponent
$\alpha=1$ (dashed line in Fig.~\ref{fig:theory}). Interestingly, the same
expression is found for late-stage coarsening in spinodal decomposition of
inter-percolating phases~\cite{Siggia1979,Tanaka2000}, but the underlying
mechanism is different. Whereas we consider an evaporation--condensation
process at intermediate time scales, the coarsening of inter-percolating
networks is a result of hydrodynamic flow on long time scales. Surprisingly,
both cases lead to the same coarsening law, Eq.~\eqref{eq:Rve}.

The extent of the viscoelastic ripening regime is limited by monomer
diffusion, with a typical diffusion time scale $\tau_{\tn D}=\bR^2/D$, and a
transition to Ostwald ripening occurs at sufficiently large cluster
size~$R_{\tn c}$~(Fig.~\ref{fig:theory}).  To determine the crossover
size~$R_{\tn c},$ we equate the two growth rates, Eqs.\ \eqref{eq:ROst}
and~\eqref{eq:Rve},
\begin{equation}
  R_{\tn{c}}^2 = \frac{16 c^{\tn{sat}}}{27 c_{\tn A}^2} \frac{D \tau_{\tn r}
    \tey}{k_{\rm B}T}\;.
\label{eq:Rcbound}
\end{equation}
Viscoelastic ripening is limited to $\bR < R_{\tn{c}}$ and will thus be
prominent if $D\tau_{\tn r}$ is large, with $\tau_{\tn r}$ the Maxwell
relaxation time. In a single-component phase, these two factors are connected
via the fluctuation--dissipation theorem. Employing the Stokes--Einstein
relation and assuming a standard situation with a dense phase~A, i.e.,
$c_{\tn A} \sim r_{\tn m}^{-3}$ with $r_{\tn m}$ the molecular size of
individual monomers, we find $R_{\tn{c}} < r_{\tn m}$, so that viscoelastic
ripening does not exist~\footnote{Viscoelastic ripening could potentially be
  found in a one-component system provided $c_{\tn A} \ll r_{\tn m}^{-3}$,
  which would occur if gas bubbles nucleate within a dense phase and grow via
  surface-tension-driven coarsening.}. We conclude that viscoelastic ripening
of precipitates or liquid droplets is expected to occur only in multicomponent
materials that can simultaneously support both fast diffusion of monomers and
sufficiently slow material relaxation. A typical example of such material is a
hydrogel with a solvent viscosity that is orders of magnitude smaller than the material viscosity.

To quantitatively delineate the different growth regimes, we consider that the
viscoelastic growth, Eq.~\eqref{eq:Rve}, is noticeable upon doubling of the
initial cluster size, which occurs on a time scale~$\tau_{\tn{ve}}$,
\begin{equation}
\tau_{\tn{ve}} = \tau_{\tn r} \frac{2\tey r_0}{\gamma} \;,
\end{equation}
whereas the transition to Ostwald ripening occurs at crossover time
scale~$\tau_{\tn c}$ determined by Eqs.\ \eqref{eq:Rve}
and~\eqref{eq:Rcbound},
\begin{equation}
\tau_{\tn c} = \tau_{\tn{ve}} \frac{4}{c_{\tn A}r_0}\sqrt{\frac{D
    c_{\tn{sat}}\tau_{\tn r}\tey}{27k_{\tn B}T}} \;.
\end{equation}
Thus, the distinct ripening regimes are characterized as: (i)~initial growth
and elastic arrest for $t < \tau_{\tn{ve}}$; (ii)~viscoelastic ripening with
growth exponent $\alpha \approx 1$ for $ \tau_{\tn{ve}} < t < \tau_{\tn c}$;
and (iii)~convergence to the LSW theory with $\alpha \approx 1/3$ for
$t > \tau_{\tn c}$.

\subsection{Power-law material response}

\begin{figure}
\centering
\includegraphics[width=0.45\textwidth]{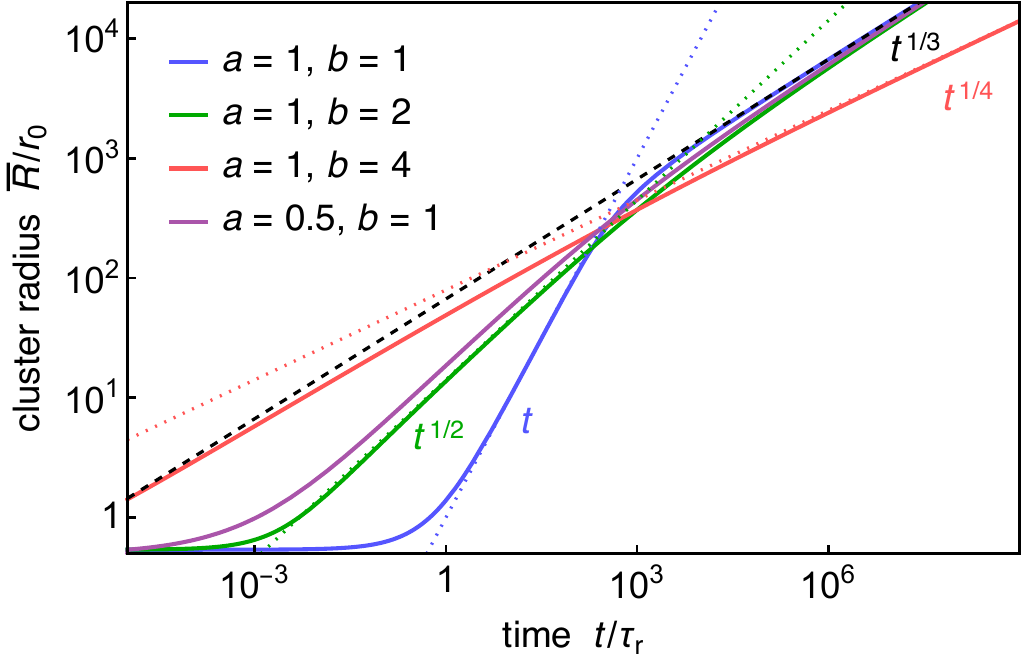}
\caption{Ripening behavior at different values of the power-law exponents in
  the strain rate, Eq.~\eqref{eq:doteps}. Numerical prediction is obtained by
  solving Eqs.\ \eqref{eq:prpt} and~\eqref{eq:doteps} (solid lines) with
  corresponding asymptotic laws for Ostwald [dashed black line,
  Eq.~\eqref{eq:ROst}] and viscoelastic [dotted lines, Eq.~\eqref{eq:Rt}]
  ripening. Prefactors:
  $C c^{\tn{sat}}_{0} = 10^{5} r_0 ^3/ (\gamma \tau_{\tn r}) $,
  $E_{\tn Y} \gg \gamma/r_0$ and
  $A \tau_r^a ( \gamma / r_0)^b = 1,100,10^7,3000$, for the blue, green, red
  and purple lines, respectively.}
\label{fig:powerlaw}
\end{figure}

The accuracy of the analytical prediction shows that our approach can be
generalized to any relaxation behavior with a strain rate that is a power-law
function of both the stress~$\sigma = p_{\tn{el}}$ and time~$t$,
\begin{equation}
  \frac{\tn{d}\varepsilon}{\tn{d} t} = A \sigma^{b} t^{a-1} \;,
\label{eq:doteps}
\end{equation}
with $A$ a materials-dependent prefactor. This general form captures power-law
fluids as well as a variety of glassy materials and
networks~\cite{Sollich1997}. Setting the exponents $a$ and $b$ both to unity
and $A$ to $(2 \tey \tau_{\tn r})^{-1}$ would correspond to the Maxwell
material analyzed above, whereas for the cytoskeleton $a=0.5$ and
$b=1$~\cite{Broedersz2010,Desprat2005}. If the elastic response is
instantaneous compared to viscoelastic relaxation, Eq.~\eqref{eq:doteps}
determines the cavity growth rate via
$\tn{d} r / \tn{d} t = r \tn{d} \varepsilon / \tn{d} t$, and together with
Eq.~\eqref{eq:prpt}, the ripening of clusters (Fig.~\ref{fig:powerlaw}).

Under viscoelasticity-limited growth, Eq.~\eqref{eq:sigmag}, and small elastic
strain, $\lambda \sim 1$, Eq.~\eqref{eq:doteps} can be analytically integrated
to obtain the general growth equation for viscoelastic ripening,
\begin{equation}
  \bR(t) = \gamma \left(\frac{Ab}{a}\right)^{1/b} t^{a/b} \;,
\label{eq:Rt}
\end{equation}
which agrees with the full numerical solution within the viscoelastic growth
regime (Fig.~\ref{fig:powerlaw}).  This regime is bounded by the minimum
cavity size~$r_0$ and the crossover to the Ostwald ripening regime. The latter
is obtained by equating the LSW and viscoelastic growth rates
[Eq.~\eqref{eq:ROst} and the time derivative of Eq.~\eqref{eq:Rt}], yielding
the crossover size
\begin{equation}
  R_{\tn c}(t) = r_{\tn m} \tilde{\tau}^{\frac{1}{3-b}} \;.
\label{eq:Rcgen}
\end{equation}
Here $r_{\tn m}$ is the individual particle size, which sets the length scale,
and $\tilde{\tau}$ the dimensionless relaxation parameter
\begin{equation}
\tilde{\tau} = \frac{8D c^{\tn {sat}} \gamma^{1-b} t^{1-a}}{27 A c_{\tn A}^2
  k_{\tn B}T r_{\tn m}^{3-b}} \;,
\label{eq:relax}
\end{equation}
that compares the diffusion rate~$D$ to the viscoelastic relaxation
factor~$A$.

\begin{figure}
\centering
\includegraphics[width=0.45\textwidth]{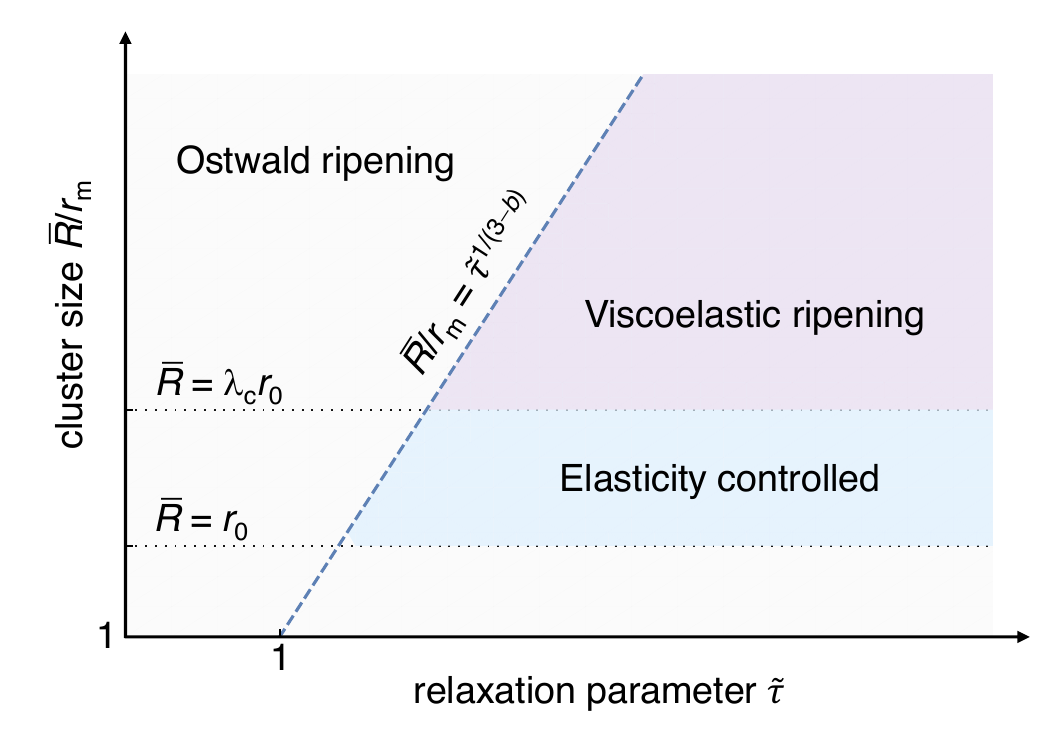}
\caption{Theoretical diagram delineating the domain growth regimes in a
  general viscoelastic material on a double-logarithmic scale: Ostwald (light
  grey), elasticity controlled (light blue) and viscoelastic ripening (light
  purple).  The boundaries are determined by the initial cavity size~$r_0$
  (lower dotted line), the reversible elastic expansion~$\lambda_c$ (upper
  dotted line) and the crossover to Ostwald ripening (dashed line) as a
  function of the relaxation parameter $\tilde{\tau}$,
  Eq.~\eqref{eq:relax}. For a Maxwell material ($a=1$ and $b=1$) the
  relaxation parameter becomes proportional to the Maxwell relaxation time,
  $\tilde{\tau} \propto \tau_{\tn r}$.}
\label{fig:PD}
\end{figure}

If the initial elastic expansion of cavities $\lambda_{\tn c}$ is significant,
$\lambda_{\tn c} \gg 1$, viscoelastic growth is preceded by a purely
elasticity-controlled regime bounded by $r_0 < \bR < \lambda_{\tn c}r_0$, that
does not depend on relaxation
properties~\cite{Dufresne2020}. Figure~\ref{fig:PD} summarizes the growth
regimes~\footnote{This figure does not cover the case in which initial
  nucleation and growth of small clusters follows a $R\sim t^2$
  scaling~\cite{RatkeVoorhees2002}.}.  Interestingly, for a strongly
shear-thinning material the behavior becomes qualitatively different from a
Maxwell material. At $b > 3$ the slope of the boundary in Fig.~\ref{fig:PD}
turns negative; small clusters are limited by Ostwald ripening, while larger
domains fall into the viscoelastic ripening regime (Fig.~\ref{fig:powerlaw}).

Having extensively analyzed viscoelastic relaxation, we note that the
viscoelastic matrix could form a brittle crack if the elastic stress exceeds
the fracture strength $p_{\tn {frac}} = \sqrt{\Ey \Gamma/R}$ and
$R \ge \Gamma/\Ey$, with $\Gamma$ the fracture energy of the
material~\cite{Kim2020}.  We find the elastic stress that can be produced by
surface tension-driven domain growth is limited to
$p_{\tn{el}} \sim \gamma/R$, Eq.~\eqref{eq:sigmag}. Thus, formation of a
brittle crack would require the surface tension to exceed the fracture energy,
$\gamma > \Gamma$. However, since surface tension is typically orders of
magnitude smaller than the fracture energy~\cite{Kim2020}, we conclude that
viscoelastic relaxation is the dominant out-of-equilibrium process leading to
stress relaxation in surface-tension driven coarsening.

\section{Conclusions}

We have addressed the problem of phase separation within a viscoelastic
material, a question of fundamental importance to different classes of soft
materials.  Through a combination of analytical calculations and Monte Carlo
simulation, we have demonstrated that such systems can exhibit a new type of
domain ripening behavior that is distinct from the standard Ostwald ripening
process.  Our analytical results, supported by MC simulations, show that
ripening of phase-separated domains typically exhibits three distinct regimes:
(i)~elasticity-controlled ripening and arrest on short time scales;
(ii)~viscoelastic ripening on intermediate time scales with a
material-dependent ripening exponent~$\alpha$, e.g., $\alpha \sim 1$ for a
Maxwell material; and (iii)~Ostwald ripening with $\alpha=1/3$ in the long
time limit. Quantitative experimental verification could be realized by
inducing phase separation in a viscoelastic gel with a known constitutive
relation and measuring the domain growth over multiple orders of magnitude
[cf.\ Figs.\ \ref{fig:theory} and~\ref{fig:powerlaw}].

Our findings provide insight into phase separation within a wide variety of
viscoelastic materials, including dense polymer solutions, gels and biological
networks like the cytoskeleton.  Whereas covalently cross-linked gels or
glassy plastics at room temperature exhibit only marginal viscoelastic
relaxation, many materials exhibit creep flow and thus domain growth is
possible on sufficiently long time scales of years or even
centuries~\cite{Capiel2020}. Thus, our findings may have implications for
predicting the aging and long-term stability of materials, which are often
determined by (micro)phase separation and domain growth, and could explain the
coarsening mechanism observed in old oil paintings~\cite{Casadio2019book}.

Interestingly, our results suggest that cells could regulate the domain size
of membrane-less organelles~\cite{Feric2016} by controlling the viscoelastic
properties of the surrounding cytoskeleton network. Conversely, measuring the
coarsening exponent could allow determination of the unknown viscoelastic
properties of cellular structures~\cite{Lee2021}. Whereas the present
work focuses on passive, surface-tension-driven phase separation,
qualitatively different ripening regimes could emerge if condensation and
growth are driven by active processes~\cite{Zwicker2017}.  Our perturbation
approach offers an avenue to explore such driven systems.

Lastly, we note that the framework presented can be applied to both fluid and
solid spheroidal inclusions, since the stress inside solid spheroidal
inclusions is constant~\cite{Eshelby1957}.  Although crystalline precipitates
are often faceted~\cite{Kim2020}, for a compact, convex inclusion the required
corrections are expected to be small~\cite{LLelasticity1959book}, so that our
framework qualitatively applies to crystalline inclusions as well. 



\section{Methods}

\subsection{Derivations}
\label{sec:derivations}
\subsubsection{Nonlinear elasticity}

The Gent phenomenological model~\cite{Gent1991,Gent1996} is valid for extension ratios
that are not too large, $\lambda \lesssim 10$, whereas for larger extension
ratios additional nonlinear effects due to finite chain extensibility come
into play~\cite{Gent1991,Crosby2016}. We model these strain-hardening effects,
which prevent an infinite expansion at $p \ge p_{\tn c}$, by adding an
additional power-law term with prefactor~$E_{\tn c}$,
\begin{equation}
  p_{\tn{el}}(\lambda) = \frac{E_{\rm Y}}{6} \left( 5- 4 \lambda^{-1} -
    \lambda^{-4} \right)  + E_{\tn c} E_{\rm Y} (\lambda-1)^\gamma \;.
\label{eq:pGent2}
\end{equation}
Any positive exponent~$\gamma$ will lead to a divergent pressure in the limit
of large expansion ratios, and the value of the exponent determines the
sharpness of the pressure--expansion relation.  We choose the lowest even
exponent, $\gamma=2$, ensuring that the response is symmetric at $\lambda=1$
and that above the cavitation pressure the cluster expansion is limited to
$\lambda_{\tn c} \sim E_{\tn c}^{-1/2}$.

The elastic energy of a spherical cavity is equal to the reversible pressure
work,
\begin{eqnarray}
E_{\tn{el}}(R,r) &=& 4 \pi \int_{r}^{R} p_{\tn{el}}(r'/r)  r'^2 \, dr' \nonumber \\ 
 &=& \frac{4 \pi r^3 E_{\tn Y}}{3} \left[ \left ( \frac{5}{6} \lambda^3 -
     \lambda^2 + \frac{1}{2}\lambda^{-1} - \frac{1}{3} \right)  \right. \nonumber \\
     &+& \left. E_{\tn c}
      \left( \frac{3}{5}\lambda^5 -\frac{3}{2}\lambda^4 + \lambda^3 -
     \frac{1}{10} \right)\right] \;.
 \label{eq:Eel}
\end{eqnarray} 
In comparison, the elastic energy of a spherical cavity in an isotropic
linearly elastic solid is~\cite{LLelasticity1959book}
\begin{equation}
  E_{\tn{el, lin}}(\lambda) =  \frac{4\pi E_{\rm Y}}{1+\nu} r^3 (\lambda-1)^2 \;.
\label{eq:Elin}
\end{equation}
In the limit $\lambda \to 1$ Eq.~\eqref{eq:Eel} reduces to
Eq.~\eqref{eq:Elin} at Poisson ratio $\nu=0.5$.

In all calculations we employ the full nonlinear elastic energy,
Eq.~\eqref{eq:Eel}. At $\lambda \sim 1$, where the linear elasticity
approximation, Eq.~\eqref{eq:Elin}, applies, our results are generally valid
for any linear elastic material at any $\nu$, provided that the prefactor
$E_{\tn Y}$ is rescaled to $3 E_{\tn Y} / [2 (1+\nu)]$.  Moreover, we assume
that there is no pinning of clusters or droplets to the cavity walls,
$p(\lambda) = 0$ for $\lambda < 1$ and, consequently,
$E_{\tn{el}}(\lambda) = 0$ for $\lambda<1$.

The equilibrium cluster size can be calculated by starting with the classical
nucleation theory (CNT), which assumes that the free energy~$F_1$ of a single
spherical cluster of radius~$R$ is composed of a surface term and a volume
term. To account for elastic deformation energy, we add an elastic
term~$E_{\tn{el}}(R,r)$, yielding the total free energy,
\begin{equation} 
F_1 = 4 \pi \gamma R^2 + 4 \pi R^3 f_{\tn v}/3  + E_{\tn{el}}(R,r)\;,
\label{eq:F1}
\end{equation}  
where $f_{\tn v}$ is the free-energy density of the condensed cluster that is
determined by the chemical potential of individual particles that constitute
the cluster. Using this expression we can compute the equilibrium cluster
size~$R$.

\subsubsection{Viscoelastic relaxation}

Since the pressure depends only on the expansion ratio~$\lambda$, the elastic
energy Eq.~\eqref{eq:Eel} is separable into a prefactor proportional to the
cavity volume~$r^3$ and a dimensionless energy density term~$W(\lambda)$ that
depends only on $\lambda$,
\begin{equation}
  E_{\tn{el}}(r, \lambda) = E_{\rm Y} r^3 W(\lambda) \;.
\label{eq:Eels}
\end{equation}
The elastic energy decreases through increase of the cavity size~$r$,
\begin{equation}
  \left(\frac{\partial E_{\tn{el}}}{\partial t}\right)_R = \left(\frac{\partial
      E_{\tn{el}}}{\partial r}\right)_{R} \left(\frac{\partial r}{\partial
      t}\right)_{R} \;.
      \label{eq:Eelchain}
\end{equation}
Using Eqs.\ \eqref{eq:Erelax} and~\eqref{eq:Eels} and noting that $R$ and $r$
are independent variables, $\partial r/\partial R=0$, we find that the cavity
growth rate depends only on the logarithmic derivative
of~$W(\lambda)=E_{\tn{el}}/(E_{\tn Y}r^3)$, Eq.~\eqref{eq:dR0dt}.  For the
extended Gent elastic model Eq.~\eqref{eq:Eel} this derivative is
\begin{widetext}
\begin{equation}
\frac{\partial \ln W(\lambda)}{\partial \ln \lambda} = \frac{5\lambda^4/6 -2\lambda^3/3-1/6+E_{\tn c}\lambda^4 (\lambda-1)^2}{5 \lambda^4/18 - \lambda^3/3-\lambda/9+1/6+E_{\tn c} \lambda(\lambda^5/5-\lambda^4/2+\lambda^3/3-1/30)} \;.
\label{eq:Alam}
\end{equation}
\end{widetext}

Equations \eqref{eq:dR0dt} and~\eqref{eq:Alam} describe the nonlinear
viscoelastic response of a strained spherical cavity.  At low strain,
$\lambda \to 1$, we can use the first-order approximation to
Eq.~\eqref{eq:Alam}, obtaining
$\frac{\partial \ln W(\lambda)}{\partial \ln \lambda} \approx 2/(\lambda-1) +
9/2$ and thus recovering the linear Maxwell material relaxation,
\begin{equation}
  \frac{\tn{d}r}{\tn{d} t} = \frac{R-r}{ \tau_{\tn r}} \;.
\label{eq:drdtlin}
\end{equation}

\subsection{Monte Carlo simulations}

The simulated system comprises $N_{\tn p}=25000$ particles suspended in an
implicit solvent and placed in a periodic cubic box of linear
size~$L=70d_{\tn{LJ}}$. The pair interaction between particles of
diameter~$d_{\tn{LJ}}$ and center-to-center distance~$d_{ij}$ is modeled via a
Lennard-Jones (LJ) potential~$U_{\tn{LJ}}(d_{ij})$
with cut-off~$2.5 d_{\tn{LJ}}$ and interaction
strength~$\epsilon_{\tn{LJ}}$. To preserve the appropriate diffusion dynamics
of individual particles, we only use local MC moves, consisting
of a random displacement of a single particle to a position within a sphere of
radius~$d_{\tn{LJ}}/2$, centered on the original position of the particle. The
simulation time scale $\tau = d_{\tn{LJ}}^2/40D$ is thus defined by the size of
the particles and their diffusion constant~$D$.

This standard LJ system is coupled to a viscoelastic background medium with
embedded spherical cavities. Every cluster containing two or more particles is
considered to be in phase~A and is subject to elastic stress, whereas isolated
particles are considered to be in phase~B and are not subject to elastic
stress.  When a dimer (a two-particle cluster) forms, a cavity with initial
radius~$r_{0}$ is created at
$\rr_{\tn{cav}}$, the center of mass of the two particles. This position
remains fixed as the cluster grows, unless the cluster fully dissolves into
individual particles, at which point the memory of the cavity location
disappears. The elastic energy of the cavity is determined by the mismatch
between the radii of the cluster~$R$ and of the
cavity~$r$. Viscoelastic relaxation effects are modeled via dissipative
expansion of cavities.  The cavity growth
rate~$\dot{r}$ is obtained from viscoelastic relaxation
[Eq.~\eqref{eq:Erelax}].  Cavities are expanded after each MC cycle of
$N_{\tn{p}}$ attempted displacement moves via $r \to r + \tau \dot{r}$.

To calculate the elastic energy we assume a dilute situation in which
individual clusters~$k$ of particles do not interact through elastic
deformations of the surrounding medium and the elastic energy of the system is
given by the sum of individual cluster contributions~$E_{\tn{el},k}$.  The
total energy of the system is thus
\begin{equation}
  E = \sum_{i,j} U_{\tn{LJ}}(d_{ij}) + \sum_{k}E_{\tn{el},k} \;,
\end{equation}
where the first sum is performed over all pairs of particles and the second
sum over all clusters~$k$.  To compute the location and position of the
clusters we use a neighbor-based cluster identification algorithm. Two
particles are considered neighbors (and thus belong to the same cluster) if
their center-to-center distance is less than $1.5d_{\tn{LJ}}$.  Each cluster
is characterized by its center-of-mass position~$\rr_{k}$ and radius of
gyration~$r_{\tn {g},k}$ that is calculated from the positions of the
individual particles in the cluster,
\begin{equation}
r_{\tn{g},k}^2 =r_{\tn{gcm},k}^2 + \frac{3}{20} d_{\tn{LJ}}^2 \;,
\label{eq:Rgi}
\end{equation} 
where the first term on the right-hand side is the radius of gyration obtained
from the centers of mass of all particles and the second term is the radius of
gyration of an individual particle. Since isolated clusters attain a spherical
shape with radius $R_{k}=r_{\tn{g},k}\sqrt{5/3}$, the coupling with the
elastic medium depends only on $r_{\tn{g},k}$ and~$\rr_k$.  The elastic energy
due to cavity expansion $E_{\tn{ex},k}$ is obtained by integrating the elastic
pressure from the unstressed cavity size $r_k$ to the expanded size $R_k$,
$E_{\tn{ex},k}=\int_{r_k}^{R_k} 4 \pi (r')^2 p_{\tn{el}}(r'/r_k)\, dr'$ for
$R_k>r_k$, and $E_{\tn{ex},k}=0$ otherwise.  To represent the displacement
$d_k=|\rr_k-\rr_{\tn{cav},k}|$ of the center of mass~$\rr_k$ of the cluster
with respect to the cavity position~$\rr_{\tn{cav},k}$, we add a linear
elastic displacement term~$E_{\tn{d},k}=\pi \Ey R_k d_k^2 $ for $R_k>r_k+d_k$
and $E_{\tn{d},k}=0$ otherwise~\cite{LLelasticity1959book}. The total elastic
energy of a specific cluster is $E_{\tn{el},k}=E_{\tn{ex},k}+E_{\tn{d},k}$,
which both suppresses the growth and immobilizes the clusters. No pinning of
clusters to cavity walls is considered, i.e., partially filled cavities are
allowed and do not provide additional free-energy contributions.

Since the aim is not to study coalescence but surface-tension-driven ripening
of spherical clusters, we restrict MC moves to single-particle moves that do
not lead to coalescence or breaking of existing clusters. Such coalescence
events can occur in principle but are very rare and negligible at dilute
conditions. Accurately incorporating coalescence would be difficult within our
simulation set-up, because at the point of coalescence the cluster shape is
highly nonspherical and thus the elastic deformation of the surrounding
medium is challenging to calculate. Therefore, we exclude the influence of
possible coalescence events, by rejecting any MC move that involves the
merging of two existing clusters or the breaking of a cluster into two
disjoint parts where at least one of these parts is not a free, unbound
particle. Our simulations thus always model surface-tension-driven ripening of
isolated spherical clusters in the dilute limit where individual clusters do
not interact directly.

Phase equilibrium is determined by the LJ interaction
parameter~$\epsilon_{\tn{LJ}}$ and the density~$\rho=N_{\tn p}/L^3$, while the
viscoelastic properties are determined by the modulus~$\Ey$, the coefficient
$E_{\tn c}$ and the viscoelastic relaxation time relative to the diffusion
time scale~$\tau_{\tn r}/\tau$. The calculations are performed at
temperature~$T=\epsilon_{\tn{LJ}}/(1.8k_{\tn B})$, which is in a two-phase
coexistence region of the LJ phase diagram (cf.\ Fig.~\ref{fig:MC}a,b).  We
set~$\Ey=k_{\tn B}T/d_{\tn{LJ}}^3$ and~$r_{0}=d_{\tn{LJ}}$, with $r_0$ the
initial cavity size corresponding to a flexible gel with crosslink distance
$r_0$~\cite{rubinstein03}. This choice permits simulations of sufficiently
large systems (i.e., containing multiple coarsening droplets) over
sufficiently long time scales, thus making it possible to calculate the growth
exponents (cf.\ Fig.~\ref{fig:MC}).  Using $r_0 \gg d_{\tn{LJ}}$ would more
accurately capture experimental systems, but would be prohibitively expensive
to simulate. We anticipate that any choice satisfying $r_0 \ge d_{\tn{LJ}}$
yields the same predictions, since the molecular details on length scales
smaller than the mesh size should not affect the condensation and coarsening
process in the viscoelastic regime.  By choosing $r_{0}=d_{\tn{LJ}}$ we miss
details of the process on length scales smaller than the mesh size, where
viscoelastic effects are not relevant.

The average cluster size~$n$ and radius~$\bR$ are calculated from all clusters
with more than five particles, to avoid counting transient small aggregates.
The mapping between simulation and theoretical parameters is obtained from
equilibrium properties of the LJ fluid and the simulation time
scale~$\tau$. We set $\gamma=3.06k_{\tn B}T d_{\tn{LJ}}^{-2}$ and
$c_{\tn A} = 1.14 d_{\tn{LJ}}^{-3}$, which are found by extrapolating LJ fluid
parameters~\cite{Janecek2006} to $\epsilon_{\tn{LJ}}=1.8k_{\tn B}T$, while
$c_0^{\tn{sat}}=4.3 \cdot 10^{-4} d_{\tn{LJ}}^{-3}$ is obtained directly from
MC simulations.

\subsection{Perturbation theory}
\label{sec:ptheory}

We consider viscoelastic effects in a first-order perturbative expansion of
the LSW Ostwald ripening theory, Eq.~\eqref{eq:ROst}.  For clarity we use dot
notation for derivatives, $\dot{R}\equiv \tn{d}R/\tn{d}t$. The growth rate of
the mean cluster size~$\dot{\bR}$ depends on the thermodynamic driving
force~$f$, the derivative of the free-energy density of $N$ spherical clusters
at constant total volume $V=\frac{4}{3}\pi N \bRRR$,
\begin{equation}
  f = \left. - \frac{\partial (F/V)}{\partial \bR} \right|_{V} \;.
  \label{eq:fbR}
\end{equation}
The growth rate can also depend on the kinetic prefactors, such as the
diffusion constant, which determine monomer transport.  Here we assume that
monomers can diffuse uninhibited through the viscoelastic medium, so that the
kinetic prefactors are constant. Thus, the first-order correction to LSW
ripening due to viscoelastic effects is
\begin{equation}
  \dot{\bR} = \dot{\bR}_{0} + \left. \frac{\partial \dot{\bR}}{\partial f}\right|_{f_{\tn{0}}}  (f-f_{\tn{0}}) \;,
\label{eq:R1app}
\end{equation} 
where $ \dot{\bR}_{0}$ and $f_{\tn{0}}$ are the growth rate and the driving
force, respectively, in the reference (LSW) system.  Within this first-order
expansion the relative cluster size distribution $h(R/\bR)$ is
time-independent and given by the LSW theory~\cite{RatkeVoorhees2002}.  We
assume that the expansion ratio~$\lambda$ has the same value for all clusters
in $h(R/\bR)$, so that the joint distribution of clusters and cavities is
\begin{equation}
\psi(R/ \bR, r/\br) = h(R/\bR) \delta(R-\lambda r)\;,
\label{eq:psi}
\end{equation} 
with $\lambda$ the sole variable and $\delta(x)$ the Dirac delta function.

To calculate the first-order correction we need to evaluate the total free
energy~$F$. The free energy of a single cluster~$F_1$ is determined by the
surface tension and elastic energy contributions [cf.\ Eq.~\eqref{eq:F1}] and
the total free energy is obtained by summing over all clusters in the system,
$F=\sum_i F_{1,i}(R_i,r_i)$. Using the mean-field approximation
$F = N C_{\psi} F_1(\bR,\br)$ in Eq.~\eqref{eq:fbR}, where $F_1(\bR,\br)$ is
the free energy of an average-sized cluster, we find the reference driving
force $f_0 = 3 C_{\psi} \gamma/(C'_{\psi}\bR^2)$ and the viscoelastic
contribution
$f-f_0 = 3C_{\psi}/(4\pi C'_{\psi}) \partial(E_{\tn{el}}/\bR^3)/\partial
\bR$. The size distribution ($\psi$) dependent prefactors
$C_{\psi}= \overline{F_1(R,r)} / F_1(\bR,\br)$ and $C'_{\psi}= \bRRR /\bR^3$
are assumed time-independent. Using these expressions in Eq.~\eqref{eq:R1app}
we obtain the first-order perturbative correction to the LSW theory
[Eq.~\eqref{eq:prpt}].

The assumption of a time-independent distribution 
does not apply in the case of elastic arrest, where $h(R/\bR)$ would gradually
change to a uniform size distribution~\cite{Schmelzer1990b}. Capturing the
time dependence in $h(R/\bR)$ would require a second-order expansion, for
which we did not find an analytically tractable expression. However, we do not
expect second-order effects to qualitatively affect the growth rate. Even in
the case of elastic arrest, the first-order theory, Eq.~\eqref{eq:prpt},
already correctly predicts an exponential approach of $\bR$ to an equilibrium
value~\cite{Schmelzer1990}.

To obtain an analytical relation between stress and cluster size at small
deformations ($\lambda \sim 1$) we consider a closed system containing a total
constant volume~$V$ of the condensed phase. Assuming that clusters can freely
exchange material via exchange of individual particles, the equilibrium number
of clusters $N$ is determined by the minimization of the total free energy of
the system, $F=N C _\psi F_1(R, r)$ [cf. Eq.~\eqref{eq:F1}].  Thus, the
equilibrium values for $N$ and~$R$ are obtained by setting
\begin{equation}
\left(\frac{\partial (N C_\psi F_1)}{\partial R}\right)_V=0 \;.
\end{equation}
Solving this equation in the limit $N\to\infty$ yields
\begin{equation} 
4\pi R^2\gamma = R\frac{\partial E_{\tn {el}}}{\partial R} - 3E_{\tn{el}}\;.
\label{eq:pEpR} 
\end{equation} 
The elastic energy can generally be expressed as a polynomial expansion,
$E_{\tn{el}}= A\varepsilon^2 + B\varepsilon^3 + C\varepsilon^4 + \ldots $,
around its zero value, $E_{\tn{el}}(\varepsilon=0)=0$, with the strain
$\varepsilon = R/r-1$. Thus, the energy scales with its derivative as
$E_{\tn {el}} \sim \varepsilon r \frac{\partial E_{\tn {el}}}{\partial R}$, so
that under small deformations, $\varepsilon \to 0$, the last term in
Eq.~\eqref{eq:pEpR} can be neglected.  Expressing the first term on the
right-hand side of Eq.~\eqref{eq:pEpR} in terms of the radial elastic stress,
$\sigma = \frac{\partial E_{\tn {el}}}{\partial R} \frac{1}{4 \pi R^2}$, we
observe that this stress is determined solely by the surface tension and the
cluster radius,
\begin{equation} 
  \sigma = \gamma / R \;.
\label{eq:sgRrel}
\end{equation}
Interestingly, $\sigma$ does not depend on the elastic properties of the
material or on the cavity size~$r$.

\begin{acknowledgments}
  We thank Peter Voorhees and Robert Style for enlightening discussions. This
  work was supported by the E.U.  Horizon 2020 program under the Marie
  Sk\l{}odowska-Curie fellowship No.~845032,  the U.S. Department of
  Energy, Office of Science, Office of Basic Energy Sciences, Division of
  Materials Sciences and Engineering, under Award Number DE-SC0020885,
  and the Center for Scientific
  Studies in the Arts at Northwestern University, which is funded by the
  Andrew W. Mellon Foundation.
\end{acknowledgments}


%

\clearpage
\begin{figure}
 \centering 
 \includegraphics[width=1\textwidth,page=1]{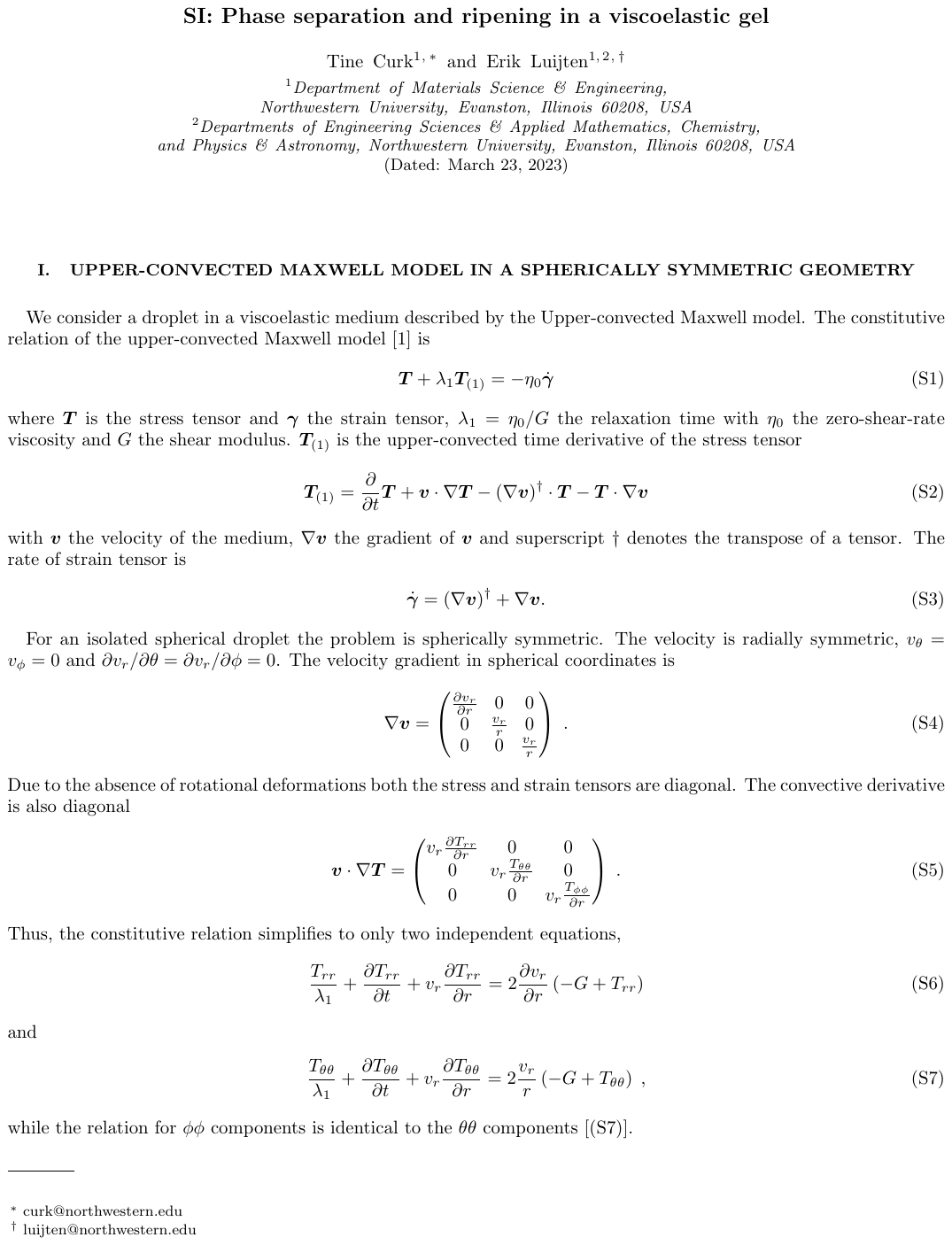}
\end{figure}
\begin{figure}
 \centering 
 \includegraphics[width=1\textwidth,page=2]{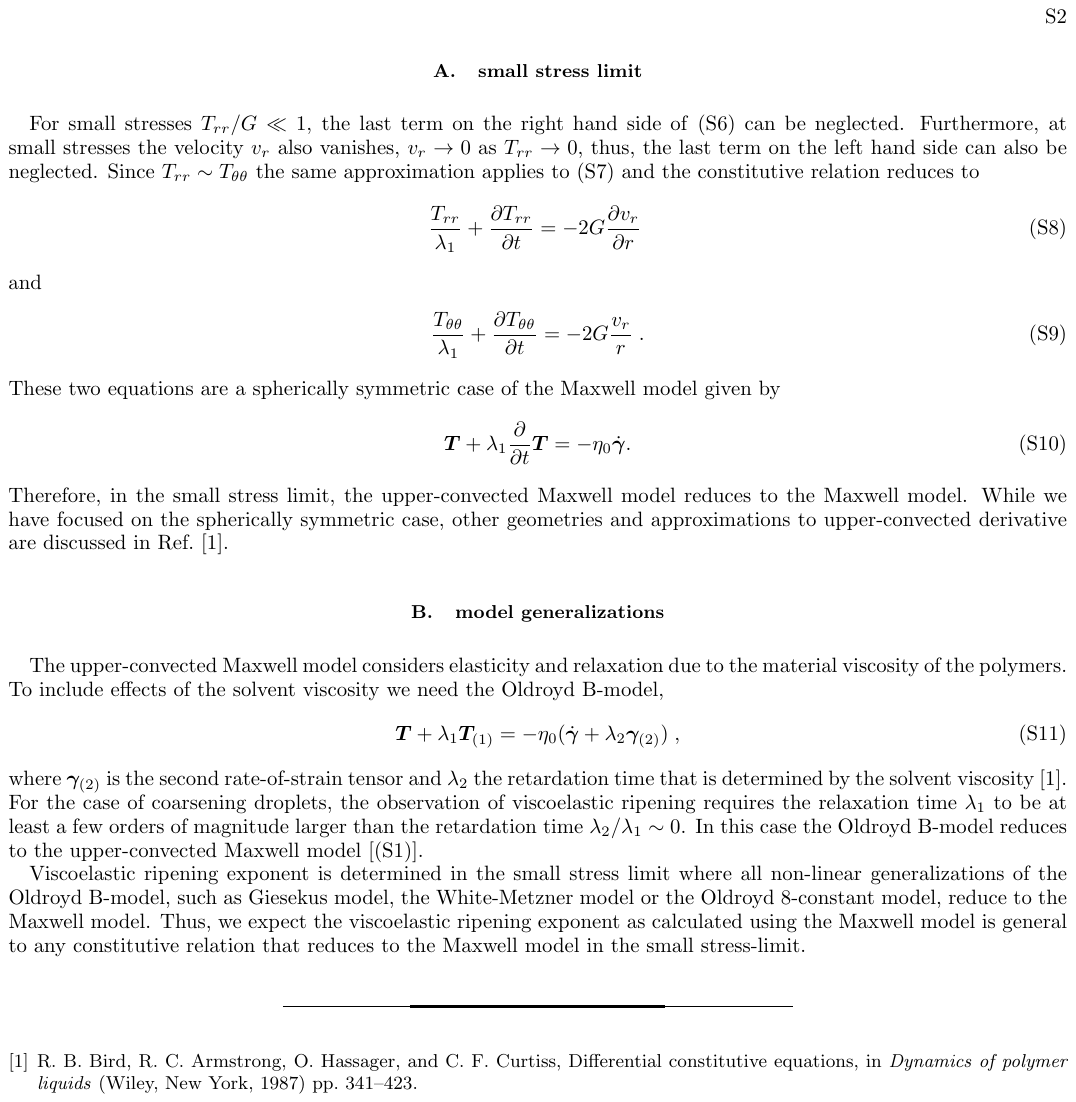}
\end{figure}

\end{document}